\documentclass[aps,prl,twocolumn,superscriptaddress,amsfonts,showpacs,floatfix]{revtex4}
\usepackage{amsmath,amssymb,bm,graphicx,epsfig,psfrag}

\newcommand\vect[1]{{\mathbf {#1}}}

\begin{document}

\title{Direct Measurement of Effective Magnetic Diffusivity  in
Turbulent Flow of  Liquid Sodium}
\author{Peter Frick}
\email{frick@icmm.ru}
\author{Vitaliy Noskov}
\author{Sergey Denisov}
\author{Rodion Stepanov}
\affiliation{Institute of Continuous Media Mechanics,  Korolyov 1,
Perm, 614013, RUSSIA}

%-------------------------------- abstract -----------------------
\begin{abstract}
The first direct measurements of effective magnetic diffusivity in turbulent flow of electro-conductive fluids (the so-called $\beta$-effect)
under magnetic Reynolds number  ${\rm Rm} >> 1$ are reported. The measurements
are performed in a nonstationary turbulent flow of liquid sodium, generated in
a closed toroidal channel. The peak level of the Reynolds number reached
${\rm Re} \approx 3 \cdot 10^6$, which corresponds to the magnetic Reynolds number
${\rm Rm} \approx 30$. The magnetic diffusivity of the liquid metal was determined by
measuring the phase shift between the induced and the applied magnetic fields.
The maximal deviation of magnetic diffusivity from its basic (laminar)
value reaches about $50\%$ .
\end{abstract}
\maketitle

%\section{Introduction}

Small-scale turbulence plays a crucial role in cosmic magnetism, providing the small-scale (turbulent)
MHD-dynamo and contributing a lot to the dynamics of large-scale magnetic fields.
The  mean field (large-scale) dynamo equations  are derived by applying the
Reynolds approach to the magnetohydrodynamics (MHD) equations, and in the framework of
the simplest case of homogeneous and isotropic (but mirror
asymmetric) turbulence they can be reduced to \cite{1966ZNatA..21..369S}
\begin{eqnarray}\label{induction}
   \frac{\partial \vect B}{\partial t}&=&\nabla \times (\vect U \times \vect B)
   + \alpha \nabla \times \vect B + \eta \bigtriangleup
   \vect B,\nonumber\\
   \nabla \cdot \vect B& =& 0,
\end{eqnarray}
where $\vect U$ and $\vect B$ describe the mean (large-scale)
velocity and magnetic fields, $\eta=(\rho\mu)^{-1}+\beta$ is the magnetic
diffusivity ($\rho$ - electrical resistivity, $\mu$ - magnetic
permeability), and $\alpha$ and $\beta$ are the turbulent transport coefficients,
describing the action of small-scale turbulent
pulsations on the mean field dynamics (see e.g. \cite{moffatt,KrauseRaedler:MeanFieldBook}).
Coefficient $\alpha$ describes the generation effects, and $\beta$ describes
the contribution of turbulence to diffusion of the large-scale
magnetic field. { The knowledge of the magnetic turbulent transport coefficients $\alpha$ and $\beta$ is
basic for astro- and geophysical applications in dynamo theory \cite{1983mfa..book.....Z}.}

Over the last decade, major efforts were directed toward the study of MHD-dynamo in laboratory experiments
(see for review \cite{Stefani2008}). The first-generation dynamo experiments are designed on the basis of strictly-specified
large-scale flow. The Riga dynamo is driven by the cylindrical screw flow \cite{2000PhRvL..84.4365G},
the Cadarache dynamo is based on a von Karman flow between two counterrotating disks \cite{2007PhRvL..98d4502M} and
even the Karlsruhe dynamo, defined as a "two-scale" dynamo, is driven by a set of strictly prescribed helical jets inside 52 tubes \cite{2001PhFl...13..561S}. In this sense, all laboratory dynamos can be classified as quasi-laminar. In spite of that,  the Reynolds numbers reached about $10^7$ and the flows were fully turbulent in all experiments. Thus, the role of turbulence is reduced in these experiments to enhancement of the diffusion of the magnetic field which,  with a constant magnetic permeability, can be considered as an increase in effective resistance of  liquid metal. The growth of resistivity can be crucial for dynamo experiments because of the corresponding reduction in  magnetic Reynolds number. However, no direct measurement of effective resistivity in dynamo facilities has been performed up to now. An indirect indication of the beta-effect has been obtained in Madison sodium facilities by comparison of measured magnetic field and magnetic field simulated on the base of measured mean velocity field \cite{2007PhRvL..98p4503S}. { An interesting scheme of eddy diffusivity estimation from hydromagnetic Taylor-Couette flow experiment, recently suggested in \cite{2009PhRvE..80d6314G}.}

The direct measurements of $\beta$ are impeded by the fact that the effect appears only under very large Reynolds numbers, when numerous side-effects prevent the accurate isolation of the $\beta$-effect. The first
attempt of such measurements was done in a flow generated by a propeller in a vessel containing liquid sodium
\cite{2001PhRvL..86.2794R}, though the authenticity of the
obtained data is questionable both with respect to the level of the
observed conductivity variations and the estimates of the
measurement errors.

A promising method of designing high Reynolds number flows (although nonstationary) in the limited mass of liquids was proposed in \cite{2002MHD....38..143F}, in which the flow was generated by the abrupt braking of a fast-rotating toroidal channel.
Installation  of diverters in the channel made it possible to create a toroidal screw flow of liquid gallium, in which, for the first time, was observed the $\alpha$-effect,  defined by a joint action of the
gradient of turbulent pulsations and large-scale vorticity
\cite{2006PhRvE..73d6310S}.
The study of the dynamics of the nonstationary flow in a torus {\it without} diverters has shown that the development of the flow in the channel is attended by a strong short-time burst of turbulent pulsations { with a peak in range on the order of $500-1000$Hz \cite{2009PhFl...21d5108N}. This burst of small-scale turbulence provides an opportunity to detect the increase in effective resistivity of liquid metal using the low frequency alternating magnetic field ($\sim 100$~Hz).} The idea of such an experiment has been realized in the nonstationary flow of liquid gallium. The toroidal channel made from textolite
made it possible to get magnetic Reynolds number less than unity \cite{2008JETPL..88..167D}.

In this paper we exploit the similar experimental scheme using a titanium toroidal channel of larger size, filled with liquid sodium, which allowed us to increase the magnetic Reynolds number by two orders of magnitude.

\begin{figure}
 \centerline{\includegraphics[width=0.43\textwidth]{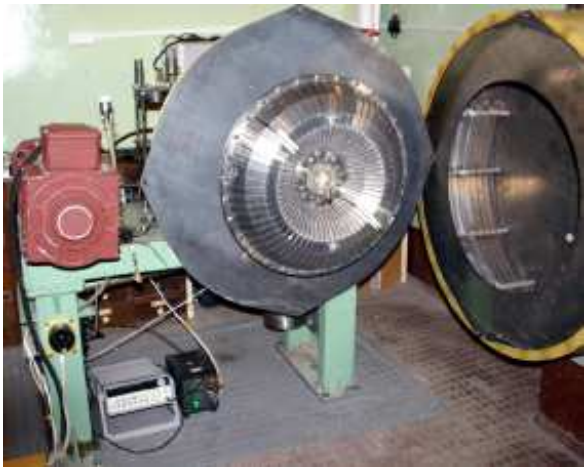}}
% \centerline{\includegraphics[width=0.43\textwidth]{Overview_bitmap.eps}}
 \caption{Titanium channel and thermostatic cover.}
 \label{fig-setup}
\end{figure}

{ The apparatus} is an electro-mechanical construction mounted on a rigid frame,
which is used as a support for a rotating  toroidal channel (Fig.~\ref{fig-setup}). % made from titanium alloy.
The torus radius is $R=0.18$\,m;  the radius of the channel cross-section is $r=0.08$\,m.
The channel was filled with sodium in the vacuum and was placed into the air thermostat.
The channel temperature may be stabilized in the range
$(50-150)^\circ$C.
The temperature sensor is mounted inside the channel and has good thermic
contact with the sodium in both liquid and solid states.

The channel is fastened  on the horizontal axis, which is also used for mounting a driving pulley,
a system of sliding contacts and a disk braking system. The frequency of the channel's rotation
is up to 45 r.p.s.~and the flow in the channel is generated by abrupt braking -- the braking
time is no more than $0.3$\,sec. The maximum  velocity of the flow is reached after channel is stopped  and achieves about $70$\,\% of the linear
velocity of the channel before braking. This means that the Reynolds number ${\rm Re}=Ur/\nu$
($\nu$ is the kinematic viscosity of the liquid sodium) reaches at maximum the value
${\rm Re}\approx 3\cdot 10^6$, which corresponds to the magnetic Reynolds number ${\rm Rm}=Ur \rho \mu \approx 30$.

\begin{figure}
\centerline{\includegraphics[width=0.43\textwidth]{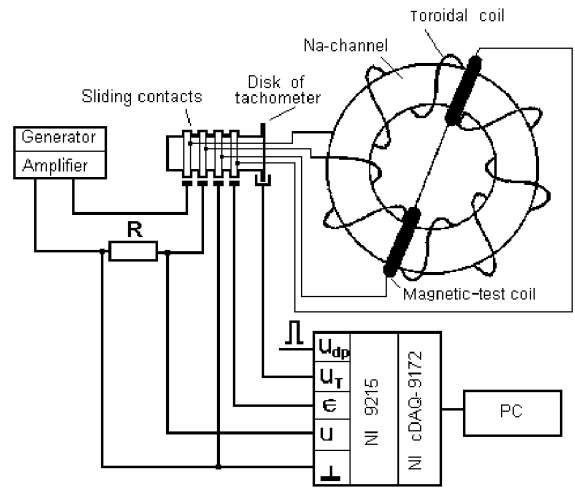}}
\caption{Schematic circuit of the measuring system. { $U_{dp}$, $U_T$, $E$ and $U$ are the driving pulse, the tachometer signal, the electromotive force of magnetic-test coil and the voltage, corresponding to applied current.}}
\label{fig-setup1}
\end{figure}

The data-gathering system is based on an NI Data Acquisition System and is a part of
an electrical measuring system, whose schematic circuit is shown in Fig.~\ref{fig-setup1}.
The 'Generator/Amplifier' block creates in the toroidal coil
a stabilized sinusoidal current with frequency $30<\nu<1000$ Hz,  which produces an alternating
toroidal magnetic field inside the channel. Besides the toroidal coil, two diametrically located magnetic-test coils are wound around the channel.

The change in phase shift $\theta$ between the measured magnetic field and the alternating current in the toroidal coil is a value,
which can be treated as a measure of logarithmic changes of diffusivity of the sodium
\begin{equation}
\Delta\theta\backsimeq C\frac{\Delta\rho}{\rho}=C\frac{\Delta\eta}{\eta},
\label{cal}
\end{equation}
where $C$ is a dimensional coefficient, which depends on the geometry and resistivity of the channel wall, and on the frequency of the applied magnetic field. The measuring system is completed with software,
based on wavelet analysis, which provides calculation of the time dependence of phase shift after signals recording. Wavelets are required
because the variation of phase shift occurs at times comparable with the
oscillation period.

\begin{figure}
\centerline{\includegraphics[width=0.43\textwidth]{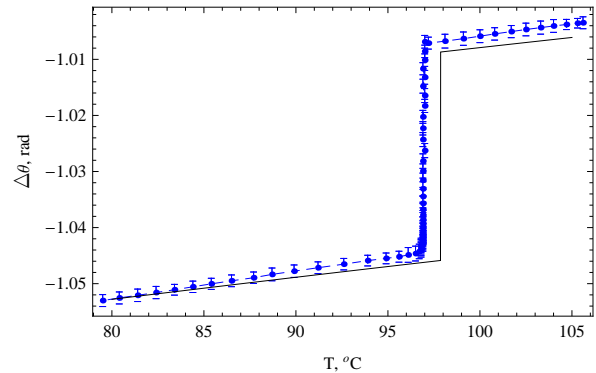}}
 \caption{Phase shift versus sodium temperature in the channel at frequency $\nu=97$\,Hz: experiment (points) and simulations (solid line).}
 \label{calibr}
\end{figure}

The measurement system has been tested and calibrated by measuring the dependence of the sodium resistivity on the temperature. The channel containing the sodium was cooled down from $105^\circ$C  to  $80^\circ$C. This range of temperature includes the sodium freezing point, which gives the best measure for calibration because the resistivity of the sodium decreased at that point by $31$\% percent, while the temperature remained constant. This excludes the influence of resistivity variation of titanium, coils, etc. Fig.~\ref{calibr} shows the results of phase shift measurements performed at frequency $\nu=97$ Hz, together with results of numerical simulations. For this frequency, the skin layer thickness of titanium is about $44$ mm (the mean thickness of titanium wall is about $10$\,mm) and the skin layer thickness of sodium is about $16$ mm.

Theoretical phase shift in the skin layer of an infinite cylindrical solenoid, which includes a titanium cylinder tube with sodium, fits the experimental points well, and allows us to define the factor of proportionality in relation (\ref{cal}) for each applied frequency. For the case $\nu=97$Hz, shown in Fig.~\ref{calibr},  $C=102\pm3$ mrad. For verification of the method,  an alternative approach of evaluating the sodium resistivity was used, based on the equivalent electrotechnical schematic of the transformer with the short-circuited secondary winding, which gave close results.

All dynamical experiments concerning the turbulent flow of liquid sodium were performed under the fixed temperature
$T=(102\pm 1)^\circ$C. The estimation of sodium heating due to energy dissipation in decaying turbulent flow at the highest rotational velocity $f=50$ r.p.s. (considering that its entire  kinetic energy will dissipate in the heat) gives $\Delta T\approx 0,8^\circ$C, which corresponds to variations of resistivity less than $0.5$\%.

\begin{figure}
\centerline{\includegraphics[width=0.43\textwidth]{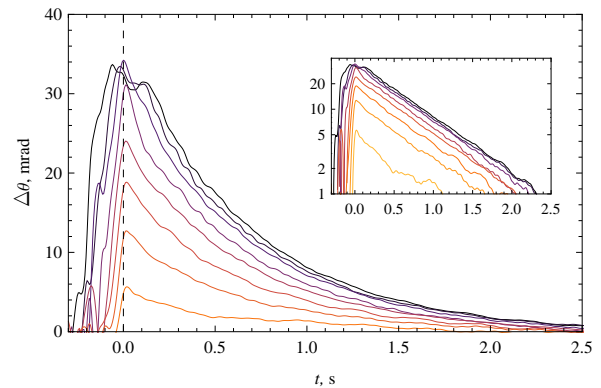}}
 \caption{Phase shift variations with flow evolution for channel rotation rate $\Omega=10, 15,..., 40, 45$\,r.p.s. (bottom-up) in linear and lin-log (inset) scales. $\nu=97$\,Hz.}
 \label{fig3}
\end{figure}

{\bf Results and Summary.} The rotational velocity $\Omega$ varied from $10$ to $45$ r.p.s. with a step of $5$ r.p.s. Measurements for all $\Omega$ were performed using three different frequencies $\nu$ ($53$, $66$ and $97$ Hz). The evolution of the phase shift, measured at frequency $\nu=97$\,Hz for different
velocity of channel rotation $\Omega$, is shown in Fig.~\ref{fig3}. Each
curve is the result of averaging over 10 realizations. The end of
braking is defined as the reference time point ($t=0$).
One can see that braking generates the turbulent flow, the maximal
intensity of which coincides with the end of braking. At this moment
the phase shift also reaches its maximum. Later on, turbulent pulsations
rapidly decay and the phase shift reduces to zero.

\begin{figure}
\centerline{\includegraphics[width=0.43\textwidth]{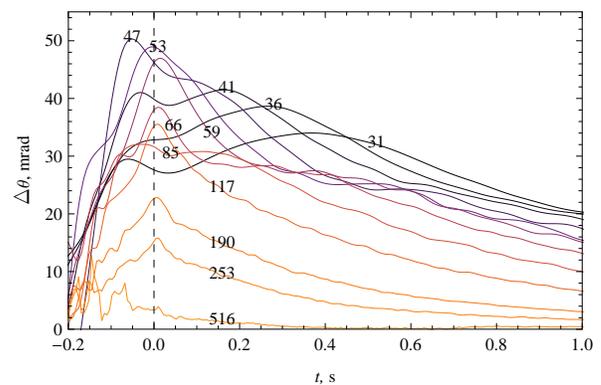}}
 \caption{Phase shift variations with flow evolution for channel rotation rate $\Omega=40$\,r.p.s. and different frequency $\nu$, shown near each curve. }
 \label{fig5}
\end{figure}

The inset of Fig.~\ref{fig3} shows that the measured phase shift decays exponentially,
which contradicts the ideas about the free decay of developed turbulence, which are rested on the power laws.
The turbulent boundary layer in the nonstationary toroidal flow is developed in a very specific way. This was found in studies of the dynamics of a similar flow of liquid gallium, which have shown
that the decay of the mean energy of the turbulent flow in the toroidal channel follows the $t^{-2}$ law, while the burst of turbulent pulsations attends the flow formation and deceases abruptly \cite{2009PhFl...21d5108N}. { This is an additional  argument to suppose that the measured phase shift is mostly caused by small-scale turbulence, but not by the dynamics of the mean flow.}

We have examined the flow across a broad range of frequencies, $31\leq \nu\leq 516$\,Hz (the skin layer thickness varies then from $29$ to $7$ mm). Fig.~\ref{fig5} shows the phase shift evolution for different frequencies and it confirms the general idea that the turbulent diffusivity should follow the intensity of turbulent pulsation, which grows from the wall of the channel to its center -- at a low frequency the contribution of the central part of the flow is larger and the $\beta$-effect is more pronounced. At its highest frequency the measuring system senses only the boundary layer, which developed first:  for $\nu=516$\,Hz the phase shift achieves the maximum at $t=-0.15$\,sec, while for $\nu=31$\,Hz the maximum appears only at $t=0.4$\,sec.

\begin{figure}
\centerline{\includegraphics[width=0.43\textwidth]{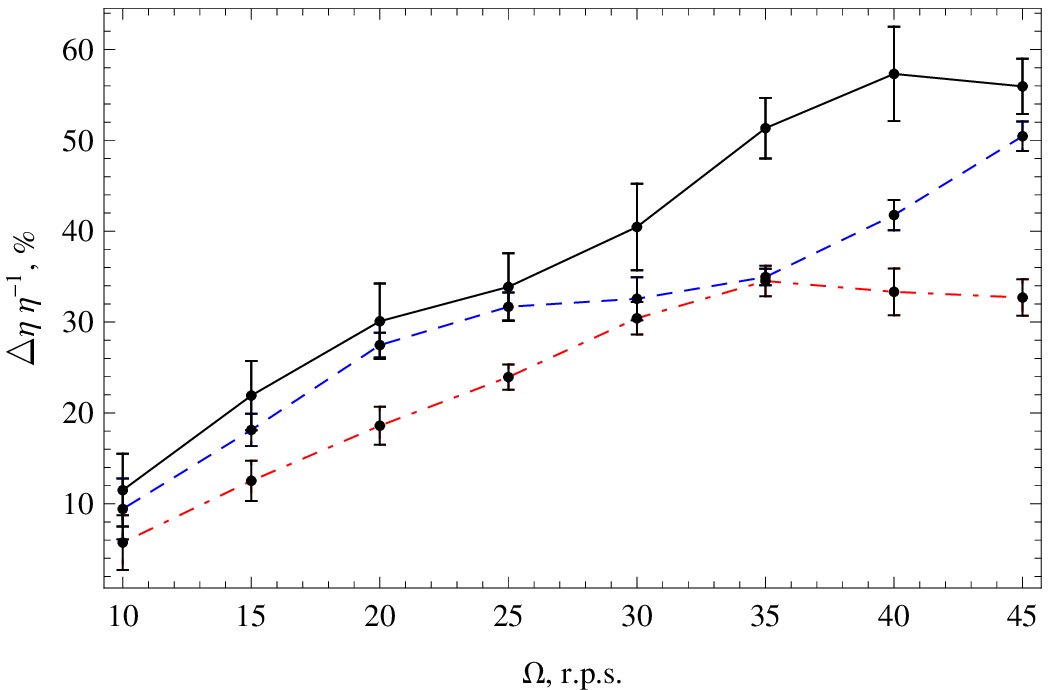}}
\centerline{\includegraphics[width=0.43\textwidth]{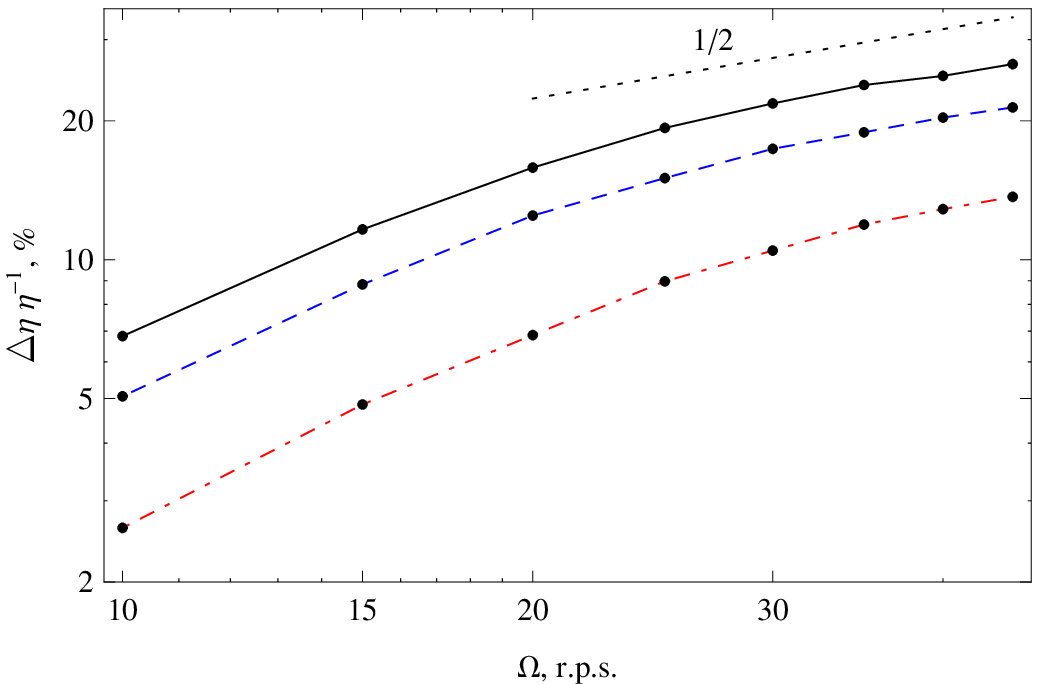}}
 \caption{Relative increase of magnetic diffusivity (percentage)  versus channel rotation rate $\Omega$ at the end of braking (top) and at $0.7$\,sec later (bottom): $\nu=53$\,Hz (solid, black), $\nu=66$\,Hz (dashed, blue), $\nu=97$\,Hz (dash-dot, red). The lower panel is shown in logarithmic scale; dots show the power law "1/2".}
 \label{fig4}
\end{figure}

In Fig.~\ref{fig4} we show how the observed $\beta$-effect depends on the intensity of the mean flow (on the Reynolds number, which is defined by the channel rotation rate before braking). First,  we show (in the upper panel) the maximal deviation of effective magnetic diffusivity, which corresponds to the end of braking, from the basic value. Measurements are taken using three frequencies: $\nu=53$, $66$, and $97$ Hz. Changing frequency, we vary the depth of penetration of magnetic field into the turbulent flow. As the frequency is lowered, the thicker the skin layer becomes and the more pronounced is the observed $\beta$-effect. The maximal value (for $\Omega=45$\,r.p.s. and $\nu=53$\,Hz) exceeds $50$\%. At low rotation rates the effect increases monotonically, in a similar manner for all frequencies; however, with $\Omega> 30$\,r.p.s., the monotony is disrupted and the curves develop in disorder.
Examining individual curves for different realizations, it is possible to see that with  high rotational speeds, the structure of the curve near the maximum becomes very complex -- the maximum becomes wider with a kind of plateau, against the background which appears to be separate distinct maximums. All these peculiarities disappear very shortly -- in Fig.~\ref{fig3} one can see that at $t\approx 0.2-0.3$, all curves evolve quite similar without any deviation. We show in the lower panel of Fig.~\ref{fig4} the deviation of effective magnetic diffusivity at $t=0.7$\,sec. Then all three curves show similar monotonic increase of the $\beta$-effect. Shown in logarithmic scales, they display a tendency toward a power law $\Delta\eta \sim \Omega^{1/2}$ at high rotational velocity.

So, the measurement of electric conductivity in the  nonstationary
fully developed turbulent (${\rm Re} \lesssim 3\cdot10^6$) flow of liquid sodium in a closed channel shows that the effective magnetic diffusivity essentially increases with the Reynolds number. For the maximal rotation rate
$\Omega=45$\,r.p.s., which corresponds to ${\rm Rm} \approx 30$,
the maximal deviation of magnetic diffusivity reaches
about  $50$\%.
Experiments with liquid gallium at low  magnetic Reynolds
number (${\rm Rm} < 1$) revealed  a quadratic  like dependence
$\beta \sim ( {\rm Rm})^2$ \cite{2008JETPL..88..167D}, which corresponds to general
conceptions of the beta-effect for low ${\rm Rm}$. Our results show that the quadratic law
does not hold at moderate ${\rm Rm}$. Note that the
turbulent viscosity in {\it stationary} pipe flows at high $\rm Re$ increases as
$\nu_t \sim {\rm Re}^{1/2}$ \cite{Schlihting}
and our results show at the highest Reynolds numbers a tendency to the same power law.
One should treat
the obtained dependence to the case of stationary pipe flow, or to homogeneous turbulence, with great caution.
However,  in view of the fact that the problem of measuring the
examined characteristic in  real flows is very complicated, and that experimental data are completely absent, measurement of the
effective magnetic diffusivity in the turbulent medium, even in
one particular case, is an important step toward the  experimental
substantiation of general MHD-dynamo conceptions.

This work was supported by ISTC project 3726 and RFBR-SNRS grant
No.~07-01-92160.

\bibliographystyle{apsrev}
\bibliography{beta2}
\end{document}